\documentstyle[multicol,epsfig,aps,prl,array]{revtex}
\newcommand{\bleq}{\ifpreprintsty
                   \else
                   \end{multicols}\widetext \vspace*{-3.5ex}{\tiny
                   
		\noindent\begin{tabular}[t]{c|}
                   \parbox{0.493\hsize}{~} \\ \hline \end{tabular}}
	                              \fi}
\newcommand{\eleq}{\ifpreprintsty
                   \else
                   {\tiny\hspace*{\fill}\begin{tabular}[t]{|c}\hline
                    \parbox{0.49\hsize}{~} \\
                    \end{tabular}}\vspace*{-2.5ex}\begin{multicols}{2}
	            \narrowtext
                    \fi}
\newcommand{\bcols}{\ifpreprintsty\else\begin{multicols}{2} 
	\narrowtext\fi}
\newcommand{\ecols}{\ifpreprintsty\else\end{multicols}\fi}

\draft
\widetext

\begin{document}
\title{Smoluchowski's equation 
for cluster exogenous growth } 
\author{St\'ephane Cueille and Cl\'ement Sire}
\address{Laboratoire de Physique Quantique (UMR C5626 du CNRS),
Universit\'e Paul Sabatier\\
31062 Toulouse Cedex, France.\\}

\date{\today}
\maketitle
\begin{abstract}
We introduce an extended Smoluchowski equation  describing coagulation 
processes for which clusters of mass $s$ grow between collisions 
with $\dot{s}=As^\beta$. A physical example, dropwise condensation
 is provided, and its collision kernel $K$ is derived. In the general
case, the gelation criterion is determined. Exact solutions are found and
scaling solutions are investigated.  Finally we show
how these results apply to  nucleation of discs on a plane.
\end{abstract}

\bcols
Aggregation processes are of practical interest in many fields of
science and technology, including chemistry, material sciences, heat
transfer engineering, atmosphere
sciences, biology, astronomy, among others \cite{friedlander77,family84}. In a generic aggregation
process, clusters of ``mass'' $s$, with mass distribution $N(s,t)$, 
can encounter collisions with other clusters. Whenever two clusters collide,
they can stick to form a new cluster with mass conservation.
 Although simple numerical models
\cite{meakin83,kolb83,meakinrev92} have proved to be a very successful tool
in the study of these phenomena, much of our theoretical understanding is still
to a large extent based on Smoluchowski's equation \cite{smoluchowski18},
 \begin{eqnarray} \label{smoleq}
\partial_t N(s,t)
&=\frac{1}{2}\int_0^s N(s_1,t)N(s-s_1,t)K(s_1,s-s_1)\,ds_1\nonumber\\
&- N(s,t) \int_0^{+\infty}\!\! N(s_1,t) K(s,s_1) \,ds_1,
\end{eqnarray} 
where the collision kernel $K(s_1,s_2)$, is the collision rate between 
clusters of mass $s_1$ and 
$s_2$, and contains the physics of the aggregation process.
 Smoluchowski's equation is obtained in mean-field,  neglecting density 
fluctuations, and  
is valid above a model-dependent upper critical
dimension $d_c$. 

Smoluchowski's equation can be used  to study the scaling properties 
of a given aggregation model.
Dynamic scaling corresponds to the fact, observed both in experiments and
 in numerical models \cite{vicsek85}, that the size
distribution is asymptotically scale-invariant at large time, 
$N(s,t)\sim S(t)^{-\theta} f(s/S(t))$, where the typical size $S(t)$
usually diverges as $t^z$. Smoluchowski's equation can be used to determine
$z$ and $\theta$, and the profile of the scaling function. 

Van Dongen and Ernst \cite{vdg85prl,vdg87scal} have extensively studied
the scaling solutions of Smoluchowski's equation. They have shown that 
collision kernels  could  be characterized by two exponents $\lambda$ and 
$\mu$, 
 \begin{eqnarray}\label{genscal1}
K(bx,by)=b^\lambda K(x,y),\\
K(x,y)\sim x^\mu y^{\lambda-\mu} \,\,\, (y \gg x), \label{genscal2}
\end{eqnarray}
which determine the small $x$ behavior of $f(x)$. If $\mu<0$, $f$ is
bell-shaped, while for $\mu \geq 0$, $f(x)\sim x^{-\tau}$. For $\mu>0$,
$\tau=1+\lambda$, while for $\mu=0$, $\tau$ is nontrivial and
$\tau<1+\lambda$. In the latter case, such a crucial quantity as the
exponent  describing the decay of the 
total number of clusters ($n(t)\propto t^{-z'}$) is given by $z'=z(2-\tau)$ 
if $\tau>1$, and thus depends on $\tau$.  The determination of $\tau$ was
very difficult until a very effective variational method was recently 
introduced \cite{nontriv97}. 
      
  Despite this reasonably satisfactory state of the art, standard 
Smoluchowski's equation does not describe all aggregation processes, and
some extensions have been provided to include for instance 
fragmentation of clusters \cite{family86}, or
  source and sink terms \cite{white82,racz85,hayakawa87}. In this Letter, we would like  to consider another
 feature, the fact that  for  certain aggregation phenomena, clusters 
do not grow solely due to collisions with other clusters,
 but also collect some mass
from the ``outside'' between collisions events. A practical example is
{\it dropwise condensation}, a daily life phenomenon (water condensation on a
mirror) which has also  important consequences
in material sciences or heat transfer engineering, which motivated most of
the early studies. Since the seminal work
of Beysens and Knobler \cite{beysens86}, the  resulting fascinating
droplets patterns, or ``breath figures'', have attracted much 
interest, and simple computer models have been introduced to study the
kinetics of droplet nucleation \cite{family88,family89,fritter91,meakindrop92},
 the asymptotic surface (or
line) coverage \cite{derrida91}, or the time evolution of the ``dry'' fraction
(the surface fraction which has never been touched by any droplet)
\cite{marcos95} (also see the review by Meakin \cite{meakindrop92}).

In dropwise condensation on a $d$-dimensional substrate, individual droplets
form $D$-dimensional hyperspherical caps, and
 grow by absorption from the
vapor phase, which leads to a growth law for the droplets radius,
$\dot{r}\propto r^\omega$, ($\omega=-2$ for water on a plane 
\cite{meakindrop92}), or, in terms of the droplet mass $s=r^D$, 
$\dot{s}=As^\beta$, with $\beta=(D+\omega-1)/D$. When two droplets 
overlap, they coalesce due to surface tension, to form a new hyperspherical
droplet, with mass conservation.   Thus, dropwise condensation can be 
understood as a simple aggregation phenomenon, where droplets grow between
collisions ({\it exogenous growth}), as well as through collisions.
Such a picture is the basis of most simplified computer models, such as 
the  droplets growth and coalescence model of Family and Meakin
\cite{family89}.

Now, what is the corresponding mean-field kinetic equation ? If there were
no collisions, the equation would simply be a continuity equation 
expressing  the conservation of the number of droplets,
\begin{equation}
\partial_t N(s,t)+\partial_s (\dot{s} N)(s,t)=0,
\end{equation} 
with $\dot{s}=s^\beta$ (in the following, we set $A=1$). Then, the collision
rate of two growing droplets of size $s_1$ and $s_2$ 
is, in mean-field, just proportional to the
time derivative of the cross section $\sigma(s_1,s_2)\propto
(s_1^{1/D}+s_2^{1/D})^d$, thus 
$K(s_1,s_2)\propto (\dot{s_1}s_1^{1/D-1}+\dot{s_2}s_2^{1/D-1})(s_1^{1/D}+
s_2^{1/D})^{d-1}$, and we obtain the following equation,
\begin{eqnarray}\label{smolgen}
&\partial_t N(s,t)
+\partial_s (s^{\beta} N)(s,t)=\nonumber \\
&\frac{1}{2} \int_{0}^{s}  N(s_1,t)N(s-s_1,t)K(s_1,s-s_1,t) ds_1
\nonumber\\
&  -N(s,t)\int_{0}^{+\infty}N(s_1,t)K(s,s_1,t) ds_1,
\end{eqnarray}
with,
 \begin{equation}\label{kernel}
     K(x,y)=(x^\frac{\omega}{D}+y^\frac{\omega}{D})(x^\frac{1}{D}
              +y^\frac{1}{D})^{d-1}.
     \end{equation}
Multiplicative constants were set to $1$ by a  rescaling of $t$.
The expression for $K(x,y)$ is in agreement with an early work of 
Vincent \cite{vincent71}, who found a Smoluchowski equation for $\omega=-2$,
$d=2$ and $D=3$. However, the left hand side of his equation is
erroneous since it does not conserve the number of particles. The kernel 
has the homogeneity $\lambda=(d+\omega-1)/D$, as predicted in
\cite{family89} from scaling arguments. Eq. (\ref{smolgen})  for a generic 
collision kernel actually describes a 
general irreversible aggregation process with the exogenous growth law
 $\dot{s}=s^\beta$,
and we would like to study the  properties of such an equation.
 
{\it Gelation criterion \--} A very important issue in aggregation problems 
is the possible occurrence of gelation. Gelation corresponds to the
formation of an infinite cluster at a {\it finite time} (in the
thermodynamic limit). Common applications are found in food industry for 
instance. Another application in material sciences is the formation of fractal
aerogels with intriguing physical properties. In terms of Smoluchowski's 
equation, gelation is the phase transition  associated to the breaking of
the mass conservation through collisions. Adapting standard arguments 
\cite{vdg87scal}, let us consider the net mass flux from clusters 
with $s\leq L$ 
towards clusters with $s>L$, $J_L(t)$. From Eq. (\ref{smolgen}), it is easily seen
that,
\begin{eqnarray}\label{massflux}
J_L(t)&=&L^{1+\beta}N(L,t)\nonumber \\
&+&\int_0^L dx\, x N(x,t)\int_{L-x}^{+\infty} dy \, K(x,y)N(y,t), 
\end{eqnarray}
Now, in the absence of an infinite cluster, the mass conservation through
collisions requires that  $J_L(t)\to 0$ when $L\to \infty$. If gelation
occurs at $t_g$, the infinite cluster collects some mass through exogenous
growth and collisions, and $J_L(t>t_g)$ has a finite $L\to \infty$ limit,
which implies that $N(s,t)$ has a slowly vanishing large $s$ tail.
 Inserting  the ansatz $N(s,t)\sim A(t)s^{-\tau}$ for $t\geq t_g$, we find 
$\tau= \mbox{max} (1+\beta, (3+\lambda)/2)$. One must have $\tau>2$, since
the total mass contained in finite clusters must be finite. This shows 
that gelation can occur only if $\lambda> 1$ or $\beta>1$.

{\it Exact solutions \--} For  Eq. (\ref{smoleq}), exact solution are known only for
 $K=1$, $K=x+y$, K=$xy$, but the existence of even few exact
 time-dependent solutions
is important to check the scaling theories. Here we shall provide two exact
solutions for Eq. (\ref{smolgen}), corresponding to $K=1$, with $\beta=0$ 
and $\beta=1$. To do this, we consider the Fourier-Laplace transform 
of $N(s,t)$, $Z(z,t)=\int_0^{+\infty} e^{-zs}N(s,t) \,ds$. 
$Z(0,t)$ is the total density  of clusters $n(t)$.
  
For $\beta=0$, the Laplace transform of Eq. (\ref{smolgen}) with $K=1$ reads,
\begin{equation}
\partial_t Z +zZ=Z^2/2- Z(0,t)Z,
\end{equation}
  With $n(0)=1$ and
$Z(z,0)=Z_0(z)$, we find  $Z(0,t)=n(t)=2/(t+2)$ and,
\begin{equation}
Z(z,t)=\frac{e^{-zt}}{(t+2)^2\left( \frac{1}{4Z_0(z)}-\frac{1}{2}\int_0^t 
\frac{e^{zt'}}{(t'+2)^2} dt'\right)}.
\end{equation}
Which leads in the scaling regime $t\to \infty$, and for a monodispersed
initial condition $N(s,t)=\delta(s-1)$ to,
\begin{equation}
N(s,t)\sim \frac{2}{t^2\ln t} e^{-\frac{s}{t\ln t}}.
\end{equation}
The total mass in the system is, $M_1(t)=1+2\ln(t+1/2)$, from
$M_1(t)=-\partial_z Z(0,t)$.

For $\beta=1$, the equation for $Z$ is,
\begin{equation}
\partial_t Z-z\partial_z Z= 
Z^2/2- Z(0,t)Z.
\end{equation}
Once again, $n(t)=Z(0,t)=2/(t+2)$, and 
the  solution is,
\begin{equation}
Z(z,t)=\frac{2}{t+2}\frac{2Z_0(ze^t)}{\left(1-Z_0(ze^t)\right)(t+2)+2Z_0(ze^t)}.
\end{equation}
For a  monodispersed initial distribution, $Z_0(z)=e^{-z}$,
$Z(z,t)$ has a pole at $z_0(t)=-e^{-t}\ln (1+2/t)$, and we can
 explicitly compute $N(s,t)$,
\begin{equation}
N(s,t)=\frac{4}{(t+2)^2 e^t} \exp\left(-s e^{-t} \ln(1+\frac{2}{t})\right),
\end{equation}
which leads in the large time limit to $N(s,t)\sim \frac{4}{t^2e^t}e^{-\frac{2s}{te^t}}$.
 
These two exact solutions support the fact that Eq. (\ref{smolgen}) exhibits
dynamic scaling, but they also show some unusual time
dependence of $S(t)$, and a more general scaling form,
\begin{equation}\label{genscal}
N(s,t) \sim Y(t)^{-1} f(s/S(t)),
\end{equation}
where $Y(t)$ does not have the hyperscaling form $Y(t)\propto S(t)^{\theta}$.

{\it Scaling theory \--} Now we want to study the scaling solutions of 
Eq. (\ref{smolgen}), with the scaling form of Eq. (\ref{genscal}), starting
from a monodispersed initial  distribution
$\delta(s-1)$. A first remark is that the distribution $N(s,t)$ is at any time strictly zero
below $s_0(t)=(1+(1-\beta)t)^{1/(1-\beta)}$.   In physical terms, this is
due
to the fact that the smallest
clusters are the one which have never collided, and thus have grown since
$t=0$ with $\dot{s}=s^\beta$. Consequently, the scaling function is strictly
zero below $x_0=\lim (s_0(t)/S(t))$. $x_0$  can be zero if $S(t)\gg
t^{1/(1-\beta)}$, but cannot be infinite. 
  
For kernels with $\lambda<1$ and $\beta<1$, 
we can have a qualitative understanding of the
scaling results to be expected. Let us examine Eq. (\ref{smolgen}). On the one hand, 
if we switch off the  collision term
({\it i.e.} the right hand side of the equation), the equation describes a set of
particles which grow in time with $\dot{s}=s^\beta$, and is associated 
with the size scale $S_g(t)\propto
t^\frac{1}{1-\beta}$.  
On the other hand, if we forget the growth term in the left hand side, we
are back with a standard Smoluchowski equation describing clustering 
with mass conservation.  The typical size
in the scaling regime is $S_c(t)\propto t^{1/(1-\lambda)}$, and
$\theta=2$ \cite{vdg85prl,vdg87scal}.

Thus, if we switch on both growth and collision, we shall observe a 
``competition'' between the dynamic size scales corresponding to both processes.
 If $\beta <\lambda$, $S_g(t)\gg S_c(t)$,
and in the scaling regime we expect $S(t)\propto S_g(t)$ and
$z=1/(1-\beta)$. If $\beta>\lambda$, on the contrary, the typical size of
particles increases essentially due to collisions, hence 
$S(t)\propto S_c(t)$ and $z=1/(1-\lambda)$. For the marginal 
case $\lambda=\beta$,  logarithmic corrections may be observed, and are
indeed present for the exact solution $K=1$, $\beta=0$.  

A detailed demonstration of the scaling results is quite intricate, since 
it involves the treatment of many subcases, and will be published elsewhere
\cite{nous}. However the main idea is quite simple. Let us define the
$\alpha$-th moment, $M_\alpha(t)= \int_0^{+\infty}
s^\alpha N(s,t)ds$. With the scaling
form  of Eq. (\ref{genscal}), the different terms of Eq. (\ref{smolgen}) read,
\begin{equation}\label{1scal}
\partial_t N(s,t) \sim 
-\frac{1}{Y}\left(\frac{\dot{Y}}{Y}f(x)+\frac{\dot{S}}{S}xf'(x)\right) ,
\end{equation}
\begin{equation}\label{2scal}
 \partial_s (s^\beta N)(s,t)\sim \frac{S^{\beta-1}}{Y} 
(x^\beta f)'(x),
\end{equation}
\begin{equation}\label{3scal}
\mbox{\rm Collision term}\sim \frac{S^{1+\lambda}}{Y^2} (...).
\end{equation}
We look for an asymptotically time independent equation consistent with the evolution equation for the total mass, obtained by
multiplying Eq. (\ref{smolgen}) by $s$ and integrating over all $s$,
\begin{equation}
\dot{M_1}=M_\beta. 
\end{equation}
 We also use the fact that $S(t)$ cannot be much smaller
than $s_0(t)$. Results for $Y(t)$ and $S(t)$, are in agreement with the qualitative picture above. 

For $\beta>\lambda$, it is found that $S(t)\propto s_0(t)$, 
thus $z=1/(1-\beta)$, and $Y(t)\propto S(t)^\theta$, with 
$\theta=2+\lambda-\beta$.
The three terms in the kinetic equation are of the
same order at large time, and,  from the remark above, the scaling function is
zero below $x_0=\lim (s_0(t)/S(t))>0$. For the droplets nucleation kernel 
with $d<D$, we have $\beta>\lambda$ and these results yield $z=D/(1-\omega)$
and $\theta=1+d/D$, in agreement with the results for the model of Family
and Meakin \cite{family89}. This scaling
theory also shows that $f(x)$ cannot diverge at small $x$, as observed for
droplets models without nucleation \cite{meakindrop92}, 
 and the fact that the scaling function is even zero below
a {\it finite} $x_0>0$ is well supported by numerical simulations \cite{nous}. 

 For $\beta\leq \lambda$, $S(t)\gg s_0(t)$, and 
the exogenous growth term (\ref{2scal}) is negligible at large time. The 
scaling equation is found to be the same as for standard Smoluchowski's
equation with the same kernel,
\begin{eqnarray}\label{scaleq}
&xf'(x)+ 2f(x)=f(x)\int_0^{+\infty} f(x_1)K(x,x_1)dx_1  \nonumber\\
&- \frac{1}{2}\int_0^x f(x_1)f(x-x_1)K(x_1,x-x_1) \,dx_1,
\end{eqnarray} 
 and the results
of \cite{vdg85prl,vdg87scal} 
can be applied. The nontrivial polydispersity exponents occurring for
$\mu=0$ kernels can be computed using the variational method introduced
recently by the present authors \cite{nontriv97}. 

For $\lambda>\beta$, 
we find $z=1/(1-\lambda)$, and $\theta=2$, the total mass in the system 
being asymptotically conserved at large time.

For the marginal case $\lambda=\beta$, we find different results for $\mu\leq 0$ kernels and
$\mu>0$ kernels. For $\mu\leq 0$ kernels, corresponding to a bell-shaped
scaling functions ($\mu<0$), or a nontrivial $\tau<1+\lambda$ ($\mu=0$),
the total mass in the system grows logarithmically in the scaling regime,
$M_1(t) \propto \ln t$, we have $S(t) \propto (t\ln t)^z$, with 
$z=1/(1-\beta)$, and $Y(t)\propto S^2(t)/M_1(t)$.   For $\mu>0$ kernels, with $\tau=1+\lambda$, there is an
additional sublogarithmic correction, leading to $M_1(t)\propto (\ln t) \ln
(\ln t)$, $S(t)\propto (tM_1(t))^z$, with $z$ unchanged, and $Y\propto
S^2/M_1$ as before. 

These results are in full 
agreement with the exact solution for $K=1$ and $\beta =0$,
 both for the asymptotics of $S$, $Y$, and
$M_1$, and for the scaling function itself. Indeed, the scaling function
 is a pure exponential, just as for the exact solution of Eq. (\ref{smoleq})
with $K=1$ \cite{smoluchowski18}.

Results for systems on the gelling boundary ($\lambda=1$ or $\beta=1$) can
also be found and will be reported elsewhere \cite{nous}. Here, we would 
like to show how this Smoluchowski equation approach can be applied to the
nucleation of discs on a plane. More generally, we consider the nucleation 
model described above, but with $d=D$. Family and Meakin \cite{family89} 
have simulated 
growth and coalescence  of droplets for $d=D=2$ and $\omega=0.5$, 
and found that in contrast
with the $d<D$ case, the scaling function  is polydispersed with a 
small $x$ divergence $f(x)\propto x^{-\tau}$ \cite{meakindrop92}. Can we 
understand this feature ? 

\begin{figure}
\begin{center}
\epsfig{figure=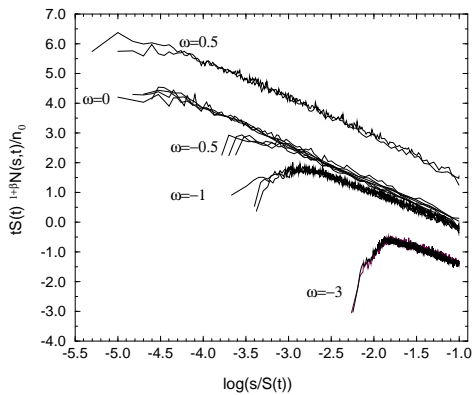,width=0.8\linewidth}
\caption{Small $x=s/S$ behavior of the  size distributions 
obtained in numerical simulations for different values of the growth
exponent $\omega$.}
\label{cross}
\end{center}
\end{figure}

For this model, the collision kernel is given by Eq. (\ref{kernel}), and we have
$\beta=\lambda + (D-d)/D$. Thus, $d<D$ corresponds to $\lambda<\beta$, 
while $d=D$ corresponds to $\lambda=\beta$, and consequently our scaling theory based 
on a Smoluchowski equation states that the scaling function vanishes below
a finite $x_0$, for $d<D$, while $f$ is bell-shaped or polydispersed for 
$d=D$, depending on $\mu$. For this kernel, we have $\mu=0$ for $\omega\geq
0$, and $\mu=\omega/D$ for $\omega<0$, which leads to the cross-over from
 a bell-shaped scaling function for $\omega<0$, to a small $x$ power
law for $\omega\geq 0$. To check this prediction, we performed simulations
of the droplets growth and coalescence model for $d=D=2$ and various values 
of $\omega$. These simulations are quite difficult since  the number of 
droplets decreases very quickly leading to
poor statistics and time range limitation. Results for the small $x$
behavior of the scaling function for different values of $\omega$  shown 
on Fig. \ref{cross} are in good agreement with the expected results. 
For $\omega=-3$, $f(x)$ has a maximum and vanishes faster than any power law
as $x\to 0$. When $\omega\to 0^{-}$, the position of the maximum of the
scaling function tends to $0$ very quickly, and the scaling function crosses
over to a power-law at $\omega=0$.  

Therefore, the mean-field scaling theory qualitatively describes the
behavior of the scaling function for $d=D$, and makes for the difference
between $d<D$ and $d=D$. Besides, for $\omega=0$, the mean-field
polydispersity exponent can be computed by the method described in
\cite{nontriv97}. We find $\tau=1.109$ \cite{nous}, which compares well with
the exponent extracted from the numerics $\tau\approx 1.2$. 

Despite this
qualitative 
success of the Smoluchowski equation  approach, it should be noticed that the upper
critical dimension of the model with $d=D$ is probably infinite, since we have 
in mean-field $M_1(t)\sim \ln t$, whereas   $M_1(t)$ is bounded for the
actual model, being proportional to the surface coverage.

In conclusion, we have generalized Smoluchowski's equation to coagulating 
systems for which clusters grow between collisions with $\dot{s}=s^\beta$. 
This equation appears to have interesting scaling properties, as found both
from exact solutions and a general study. This approach recovers scaling 
results for droplets nucleation models with $d<D$, and qualitatively 
predicts a transition in the shape of the scaling function when varying 
$\beta$,  in agreement with numerical simulations, despite mean-field 
limitations.   It would be
interesting to see if this kind of approach could  also be used to 
describe dropwise condensation with renucleation in empty spaces
\cite{family89,meakindrop92}, for which nontrivial
polydispersity exponents appear.      

We are very grateful to M. Albrecht for a critical reading of the
manuscript.

\bibliography{/h1/cueille/tex/these/these}
\ecols
\end{document}